\documentclass[12pt]{article}
\usepackage[dvips]{graphics}
\begin{document}
\renewcommand{\baselinestretch}{1.1}
\def\Fml{\phi ^{m}(l)}                  \def\Fsl{\phi ^{s}(l)}                
\def\Fsel{\phi^{s}_{exp}(l)}            \def\Fgl{\phi^{g}(l)}
\def\Ful{\phi^{u}(l)}                   \def\Fgel{\phi^{g}_{exp}(l)}
\def\Fuel{\phi^{u}_{exp}(l)}            \def\bea{\begin{eqnarray}}
\def\eea{\end{eqnarray}}                \def\FumLl{\phi^u_{mL}(l)}
\def\FumL1{\phi^{u}_{mL}(1)}            \def\fumLl{\varphi^{u}_{mL}(l)}
\def\FsLl{\phi^{s}_{L}(l)}              \def\FsL1{\phi^{s}_{L}(1)}
\def\fuLl{\varphi^u_L(l)}               \def\fuL1{\varphi^u_L(1)}
\def\fuLx{\varphi^u_L(x)}               \def\Pmpx{{\cal P}_{mpx}}
\def\FumpLl{\phi^u_{mpL}(l)}            \def\FumpL1{\phi^u_{mpL}(1)}
\def\Dmp{{\cal D}_{mp}}                 \def\fuel{\varphi^{u}_{exp}(l)} 
\def\FuLl{\phi^u_L(l)}                  \def\FXYl{\phi_{XY}(l)}
\def\FdXYl{\phi^{\,\neq}_{XY}(l)}       \def\ig{\includegraphics}

\title{\Large \bf Cosmic crystallography in a circle \\ }
\author{A.F.F. Teixeira \thanks{teixeira@cbpf.br} \\ 
{\small Centro Brasileiro de Pesquisas F\'{\i}sicas }\\
{\small 22290-180 Rio de Janeiro-RJ, Brazil} \\
       }
\date{\today}
\maketitle
\begin{abstract}
In a circle (an $S^1$) with circumference 1 assume $m$ objects distributed 
pseudo-randomly. 
In the univeral covering manifold $R^1$ assume the objects replicated
accordingly, and take an interval $L>1$. 
In this interval, make the normalized histogram of the pair separations  
which are not an integer. 
The theoretical (expected) such histogram is obtained in this report, 
as well as its difference to a similar histogram for non-replicated objects. 
The whole study is of interest for the cosmic crystallography.
\end{abstract}
\section{Introduction}
\setcounter{equation}{0}
Cosmic crystallography (CC) is a method to unveil the topology of the 
universe, and initially looked for spikes in a pair separation histogram 
(PSH) \cite{lelalu}.
Since spikes are absent in hyperbolic spaces, it appeared that the method 
was useless in such spaces. 
However, it was soon shown that not only a Clifford translation (responsible 
for a spike) press its fingerprint on a PSH, but also the other isometries 
of the space \cite{spikes1}.

When spikes are absent, the PSH of a ball containing repeated images -- the 
$\Fml$ -- is very similar to that of a ball with same radius and same 
geometry, but without duplication of images -- the $\Fsl$.
A suggestion was then made, of studying the difference of the $m$ultiply 
and the $s$imply connected histograms, $\Fml-\Fsl$ \cite{fagaus}.

To improve the method, $exp$ected functions $\Fsel$ were derived to replace 
the histograms $\Fsl$ obtained from computer simulations, for all three 
geometries with constant curvature \cite{bt1}. 
Graphs of $\Fml-\Fsel$ were obtained, clearly evincing the topology 
of an euclidian, an elliptic, and a hyperbolic three-space \cite{tsct}. 
The contribution of each individual isometry $g$ to a PSH was examined, 
and normalized histograms $\Fgl$ (defined in ref.\cite{spikes1}) were 
obtained from computer simulations \cite{tsct}; 
these simulations also gave histograms of $\Ful-\Fsel$, a previously 
unsuspected quantity \cite{v3}.

Recently the exact (noiseless) functions $\Fgel$ were given 
for the euclidian isometries \cite{bt2}. 
In the present report we finally have a first acquaintance with  
functions $\Fuel$, the exact (noiseless) counterparts 
of the 'uncorrelated' normalized histograms $\Ful$ defined in \cite{v3}. 
We examine a one-dimensional system: a universe with topology $S^{1}$, 
a circle with circumference 1; we assume the horizon at a distance $L/2$ 
on each side of an observer, so the visible universe has total length $L$; 
clearly if $L>1$ then there are repeated images in this visible universe. 
In section 2 we give a detailed description of how to obtain the expected 
uncorrelated signature $\varphi^u_{L\,exp}(l)$ when $1<L<2$. 
In section 3 we exhibit the generalization for arbitrary horizon $L/2$. 
In the Conclusion we make a few comments, and in four Appendices we 
derive a few somehow lengthy mathematical results stated in the report. 

\section{When $1<L<2$}  
In a computer simulation, we usually execute the following set 
of prescriptions to obtain the uncorrelated signature $\fuLl$: 

\vskip3mm           
\hskip3cm\scalebox{0.5}{\ig{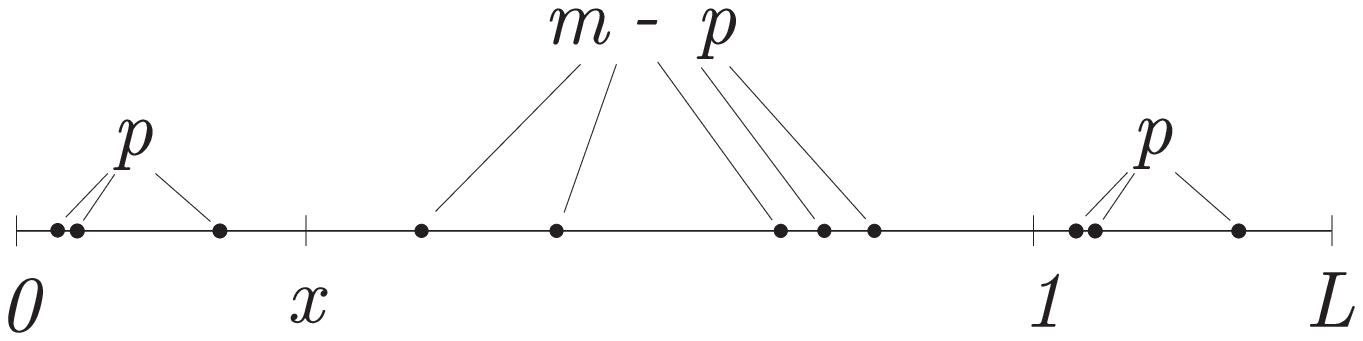}}                           
\vskip3mm     
                                          
\noindent {\small {\bf Figure 1} The distribution of objects in the interval 
$(1, L)$ is an exact copy of the distribution in $(0, x)$; here $p=3$ and 
$m=8$. } 
\vspace{5mm}
\begin{enumerate}
\item in an interval (0, 1) randomly distribute $m$ objects; 
see figure 1; 
\item in the side interval $(1, L)$ make an exact replica of the $p$ objects 
laying in $(0,x)$;  
\item measure the $(m+p)(m+p-1)/2$ separations $l$ between the total $m+p$ 
objects, and discard the $p$ correlated separations (those which have $l=1$ 
exactly); 
\item  make a normalized histogram of the 
\bea                                                          \label{1}
\Dmp = \frac{1}{2}(m+p)(m+p-1) - p \hskip1cm (1<L<2)
\eea 
uncorrelated separations;  
\item  make a large number of new normalized histograms, by repeating the 
steps 1 to 4 with same $m$ (although $p$ usually varies);  
\item take the mean of these histograms, $<\FumLl>$, and construct 
the quantity 
\bea                                                          \label{2}
<\fumLl>=(n-1-\sum_{g\in \tilde{\Gamma}}\nu_g) \Bigl[<\FumLl> -\FsLl\Bigr] , 
\eea
where 
\bea                                                          \label{3}
\FsLl=\frac{2}{L}(1-\frac{l}{L}) , \hskip5mm 0<l<L , 
\eea
and where the factor $n-1-\sum\nu_g=(m-1)L-x(1-x)/L$ is explained 
in the appendix 1;  
\item the (computer simulated) $u$ncorrelated signature $<\fuLl>$ is the 
quantity $<\fumLl>$ when $m \rightarrow \infty $; 
in practice $m>50$ usually suffices. See figure 2. 
\end{enumerate}

\vskip3mm
\scalebox{0.9}{\ig{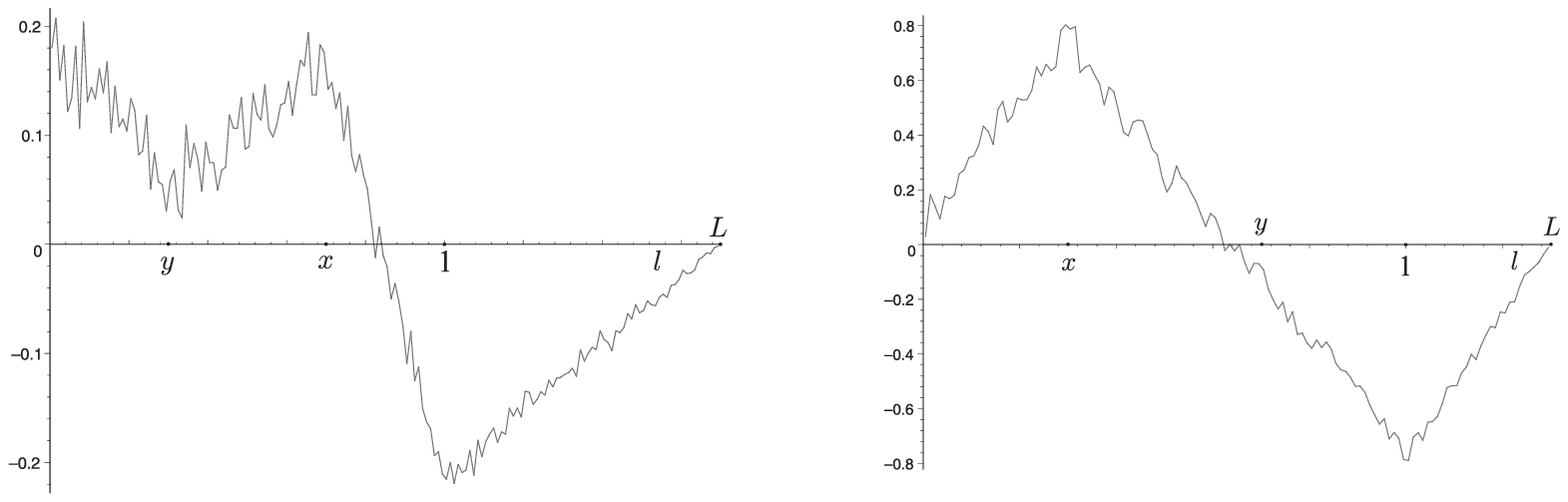}}                                  
\vskip3mm   
    
\noindent {\small {\bf Figure 2} Computer simulated functions $<\fumLl>$ 
for $\{m=2, L=1.7 \}$ and $\{m=30, L=1.3 \}$. } 

\vskip5mm
We now develop an analytical method to obtain the uncorrelated 
signature $\fuLl$. 
We are dropping the subscript $exp$ in all expected (theoretic,  
analytic, mean) probability distributions. 
Initially define the lengths $x$ and $y$ (see figure 1) 
\bea                                                          \label{4}
x=L-1 , \hskip5mm y=1-x \hskip5mm (1<L<2), 
\eea
and assume that $m$ objects are randomly distributed in (0, 1); 
the probability that $p$ objects be in the interval $(0, x)$ and 
$m-p$ objects be in the interval $(x, 1)$ clearly is 
\bea                                                          \label{5}
\Pmpx=C_m^p x^p y^{m-p} , \hskip5mm C_m^p=\frac{m!}{p! (m-p)!} ; 
\eea
irrespective of the values of $m$ and $x$ we have 
\bea                                                         \label{6}
\sum_{p=0}^m \Pmpx =1 . 
\eea

We denote as $\FumpLl dl$ the probability of finding in $(0, L)$ an 
uncorrelated pair with separation between $l$ and $l+dl$, when there are 
$m$ objects in (0, 1) and $p$ objects in $(0, x)$; clearly it satisfies 
\bea                                                         \label{7}
\int_0^L \FumpLl dl =1. 
\eea
Recall that a pair $(P, Q)$ is said $g$-correlated when the isometry $g$ 
brings one of the members to the other; the pair is uncorrelated when no 
such $g$ exists. To investigate $\FumpLl$ when $1<L<2$ we first call $A$ 
the interval $(0, x)$, call $B=(x, 1)$, and call $C=(1, L)$, 
and note that there are 
\newpage
\begin{itemize}
\item $w_{AA}=p(p-1)/2$ pairs with both members in $A$; 
\item $w_{AB}=p(m-p)$ pairs with a member in $A$ and the other in $B$; 
\item $w_{AC}=p(p-1)$ uncorrelated pairs, with a member in $A$ and 
                                             the other in $C$;  
\item $w_{BB}=(m-p)(m-p-1)/2$ pairs with both members in $B$; 
\item $w_{BC}(=w_{AB})$ pairs with a member in $B$ and the other in $C$;  
\item $w_{CC}(=w_{AA})$ pairs with both members in $C$.   
\end{itemize}
In total, there are $\Dmp$ (eq.(\ref{1})) pair separations to be considered. 

A short reflection gives that the density $\FumpLl$ can be decomposed as 
\bea                                                        \label{8}
\FumpLl=\frac{1}{\Dmp}\,\Bigl[w_{AA}\phi_{AA}(l) &+& w_{AB}\phi_{AB}(l)+ 
                         w_{AC}\phi_{AC}(l) + \nonumber \\ 
    w_{BB}\phi_{BB}(l) &+& w_{BC}\phi_{BC}(l)+w_{CC}\phi_{CC}(l)\Bigr]\, ,
\eea 
where each $\phi_{XY}(l)$ is the probability density of finding an 
uncorrelated pair of objects separated by $l$, one in $X$ and the 
other in $Y$; clearly all obey 
\bea                                                        \label{9}
\int_0^L \phi_{XY}(l) dl=1. 
\eea 

There are two basic types of $\phi_{XY}(l)$, according as $X=Y$ or 
$X\neq Y$. When $X=Y$, suppose a segment of length $\mu$, and randomly 
select two points of it; the probability that their separation lie 
between $l$ and $l+dl$ is $\phi^s_{\mu}(l) dl$ with (see figure 3) 

\bea                                                         \label{10}
\phi^s_{\mu}(l)=\frac{2}{\mu}(1-\frac{l}{\mu}) , \hskip5mm 0<l<\mu.
\eea 

\vskip3mm
\hskip4cm\ig{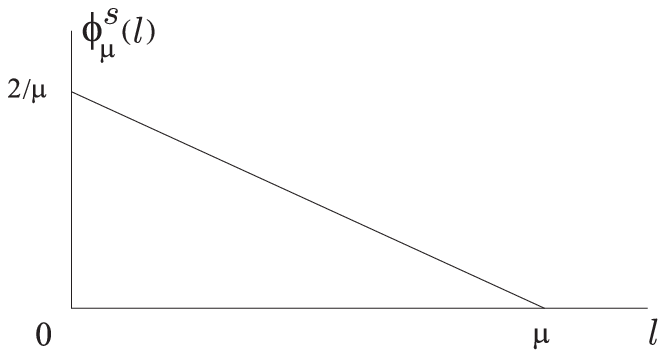}                                      
\vskip3mm   
                                       
\noindent {\small {\bf Figure 3} Pair separation density function for an 
interval $\mu$. The underlying area is 1. } 

\vspace{5mm}
When $X\neq Y$, consider two intervals with lengths $\alpha$ and $\beta$, 
with separation $\delta$ (see figure 4); 
randomly select one point in each $\alpha$ and $\beta$; 
the probability that the separation between these points 
lie between $l$ and $l+dl$ is $\phi_{\delta(\alpha\,\beta)}(l)dl$, with 
the density $\phi_{\delta(\alpha\,\beta)}(l)$ as depicted in figure 5.

\vskip3mm
\scalebox{0.8}{\ig{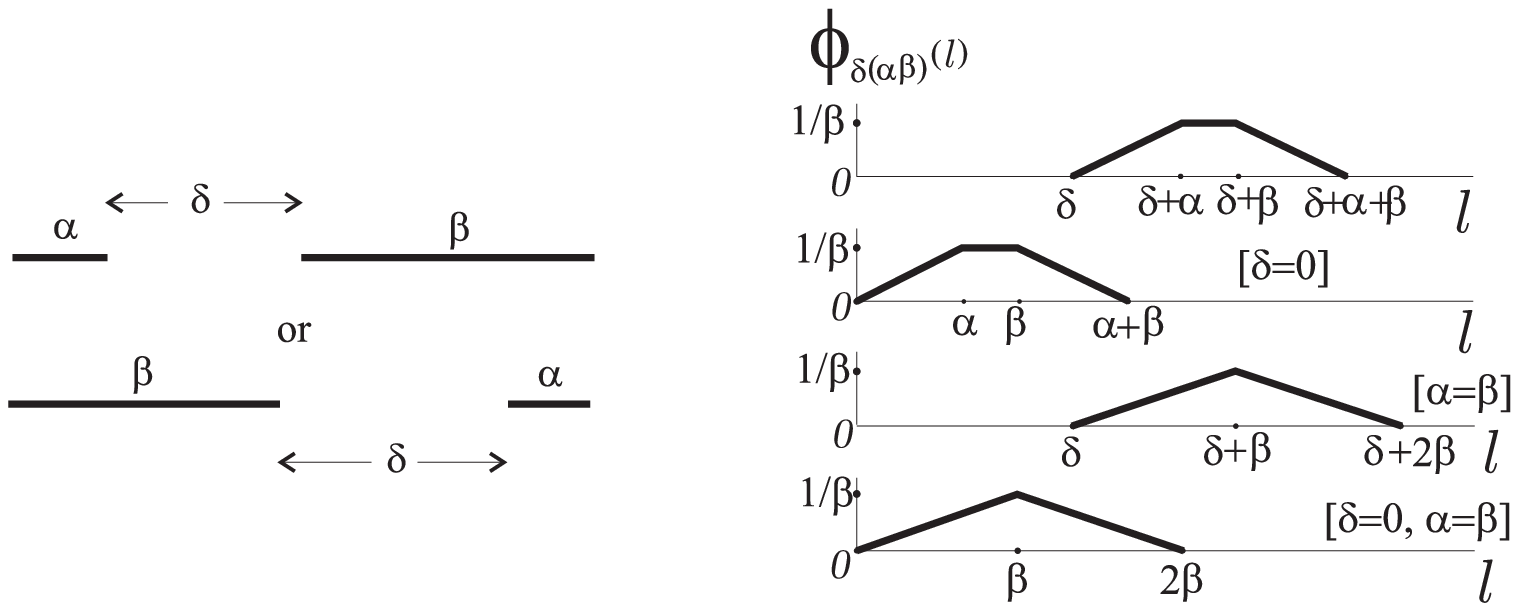}}                       
\vskip3mm  
   
\noindent {\small {\bf Figure 4} Intervals with lengths $\alpha$ 
and $\beta$, with separation $\delta$; assume $\alpha\leq\beta$. } 

\noindent {\small {\bf Figure 5} The probability density 
$\phi_{\delta(\alpha\beta)}(l)$ for $\alpha\leq\beta$ (see figure 4); 
three particular cases are also displayed; all underlying areas are $=1$. } 

\vspace{5mm}
The functions $\phi_{XY}(l)$ appearing in eq.(\ref{8}) are as displayed 
in the figure 6, for the case with $x\leq y$; for $x\geq y$ a similar set 
has to be constructed, see figure 7. 

\vskip8mm
\scalebox{0.8}{\ig{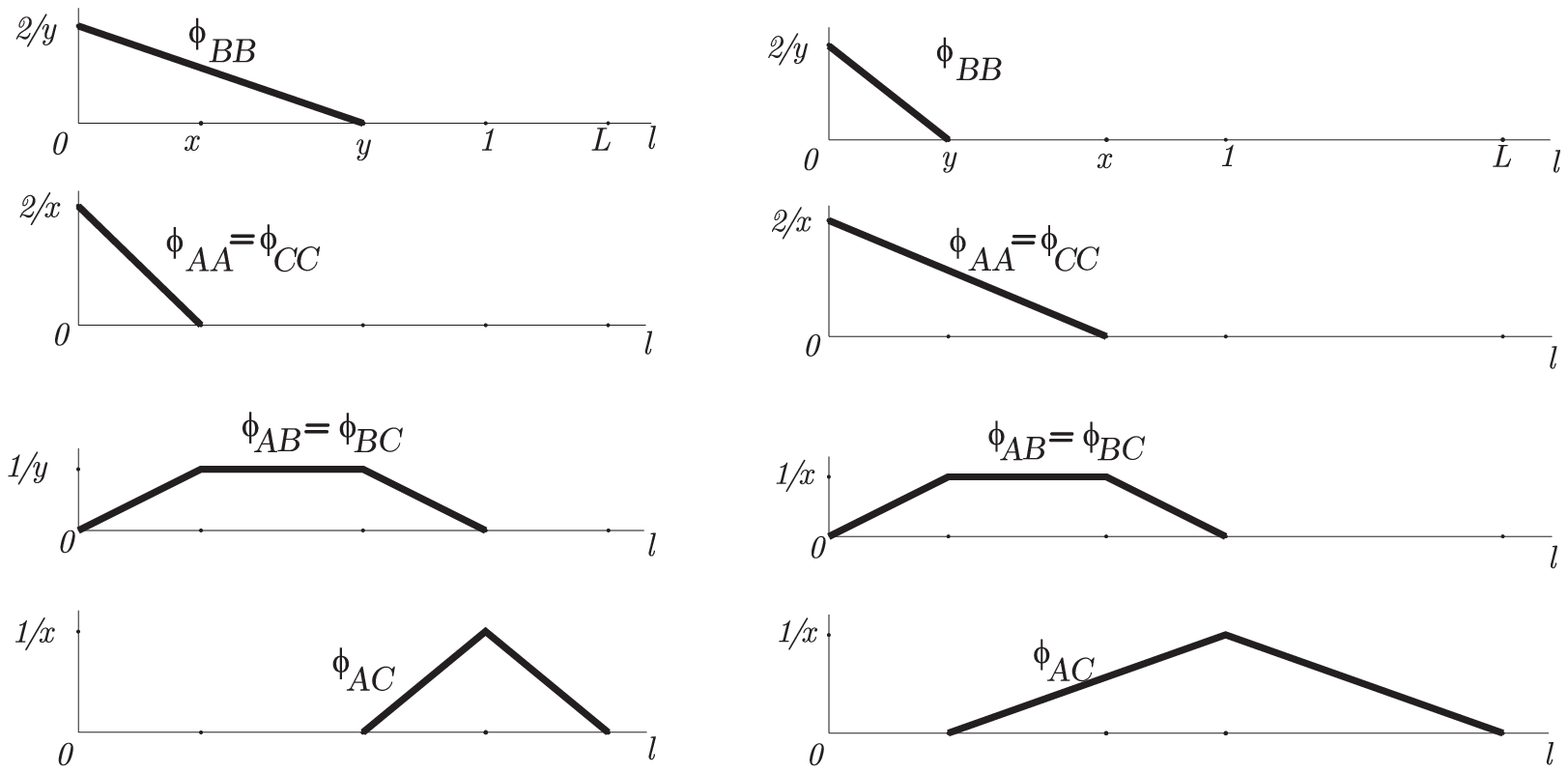}}                     
\vskip3mm
     
\noindent {\small {\bf Figure 6} The normalized functions $\phi_{XY}(l)$ 
when $1<L<2$ and $x\leq 0.5$ . } 

\noindent {\small {\bf Figure 7} The same functions when $x\geq 0.5$ . } 

\vspace{5mm}
\noindent
When $x\leq 0.5$ the density $\FumpLl$, eq.(\ref{8}), is a sequence of 
four straight segments with endpoints at $l=0, x, y, 1$, and $L$ (in this 
order), and values 
\bea                                                        \label{11}
\phi^u_{mpL}(0)&=&\frac{1}{\Dmp}[w_{BB}\frac{2}{y} + 
2w_{AA}\frac{2}{x}], \\ \nonumber  
\phi^u_{mpL}(x)&=&\frac{1}{\Dmp}[w_{BB}\frac{2(y-x)}{y^2} + 
2w_{AB}\frac{1}{x}]  \hskip5mm (x\leq 0.5), \\ \nonumber 
\phi^u_{mpL}(y)&=&\frac{1}{\Dmp}[2w_{AB}\frac{1}{y}] 
\hskip5mm (x\leq 0.5), \\ \nonumber
\phi^u_{mpL}(1)&=&\frac{1}{\Dmp}[w_{AC}\frac{1}{x}], \hskip5mm 
                                                     \phi^u_{mpL}(L)=0.
\eea

When $x\geq 0.5$ the sequence of endpoints changes to $l=0, y, x, 1$, 
and $L$, and the values of $\FumpLl$ at $l=y$ and $l=x$ become  
\bea                                                        \label{12}
\phi^u_{mpL}(y)&=&\frac{1}{\Dmp}[2w_{AA}\frac{2(x-y)}{x^2} + 
2w_{AB}\frac{1}{x}]  \hskip5mm (x\geq 0.5 )\\ \nonumber 
\phi^u_{mpL}(x)&=&\frac{1}{\Dmp}[2w_{AB}\frac{1}{x} + 
w_{AC}\frac{x-y}{x^2}]  \hskip5mm (x\geq 0.5). 
\eea   
Two examples of functions $\FumpLl$ for $1<L<2$ are shown in figure 8.

\vskip3mm 
\scalebox{0.7}{\ig{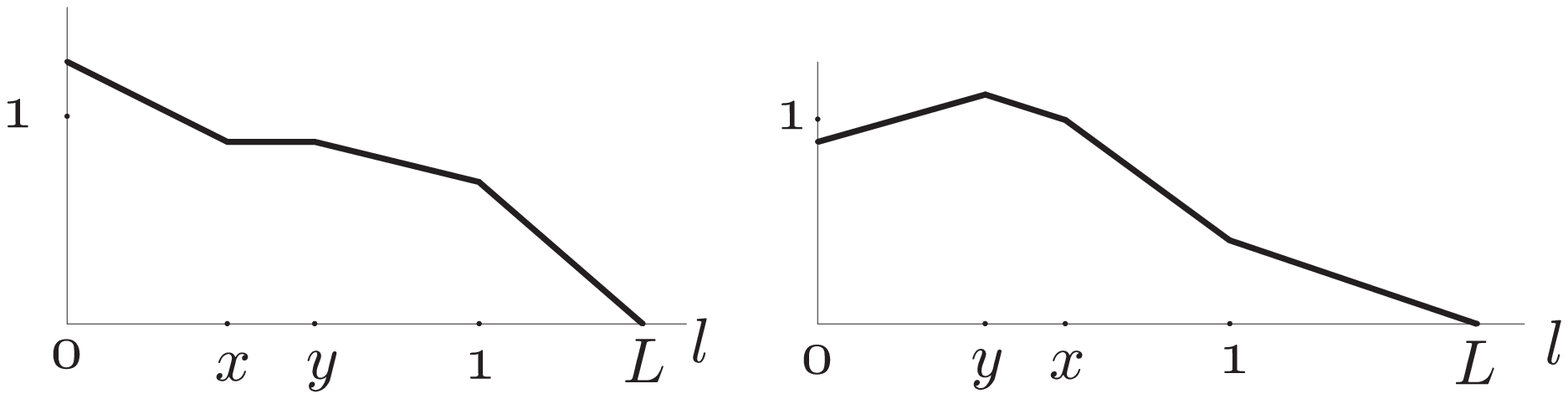}}                                
\vskip3mm
     
\noindent {\small {\bf Figure 8} The probability density 
$\phi_{mpL}^{u}(l)$ for $m=3, p=2$, and two values of $L$: 1.4 and 1.6 . 
Both underlying areas are 1. } 

\vspace{5mm}
\noindent Having the $m+1$ functions $\FumpLl$, $p=0,...,m$, we 
introduce the probability density 
\bea                                                 \label{13} 
\FumLl=\sum_{p=0}^m \Pmpx\FumpLl , 
\eea 
whose interpretation is obvious: $\FumLl dl$ is the probability that two 
uncorrelated objects randomly selected in $L$ have separation between 
$l$ and $l+dl$, when $m$ objects were randomly distributed in the 
interval (0, 1). 
Examples of $\FumLl$ are given in figure 9.

\vskip3mm
\hskip2cm\scalebox{0.7}{\ig{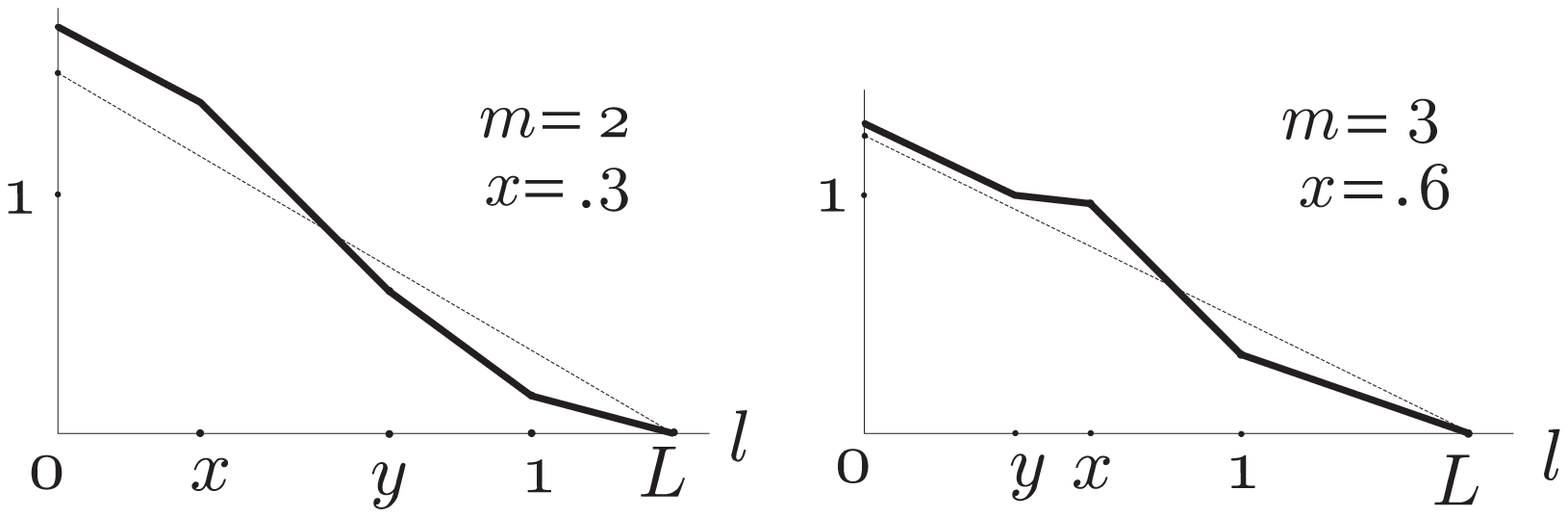}}                       
\vskip3mm
              
\noindent {\small {\bf Figure 9} Probability densities $\FumLl$ for 
$1<L<2$. The graph of $\FsLl$ is given in dotted line, for comparison. } 
 
\vspace{5mm}
Cosmic crystallography is mostly interested in systems with $m>>1$. 
In this limit the function $\FumLl$ closely resembles the simple 
triangular function $\FsLl$ (eq.(\ref{3}), fig. 3), so one is led 
to define the difference 
\bea                                                  \label{14}   
\fuLl=\lim_{m\rightarrow \infty} mL\Bigl[\FumLl-\FsLl\Bigr] , 
\eea 
the asymptotic uncorrelated signature of $L$.

We soon find that the function $\fuLl$ has a number of symmetries: 
\bea                                                 \label{15}    
\varphi_L^u(0)=\varphi_L^u(L/2)=\varphi_L^u(L)=0, \hskip5mm 
\varphi_L^u(x)=-\varphi_L^u(1). 
\eea
In other words, every $\varphi_L^u(l)$ with $1<L<2$ is composed of three 
line segments, with the first segment parallel to the third 
(see figure 10). 
As expected, the entire graph of $\fuLl$ is uniquely fixed by the 
number $f(L)$, the value of $\fuLl$ at $l=x$; 
in the appendix 2 we show that 
\bea                                                  \label{16} 
f(L)=\frac{8xy}{L^3}   \hskip5mm (1<L<2). 
\eea 
A plot of $f(L)$ valid for arbitrary $L>1$ is given in figure 11.

\vskip3cm 
\hskip1cm\scalebox{0.8}{\ig{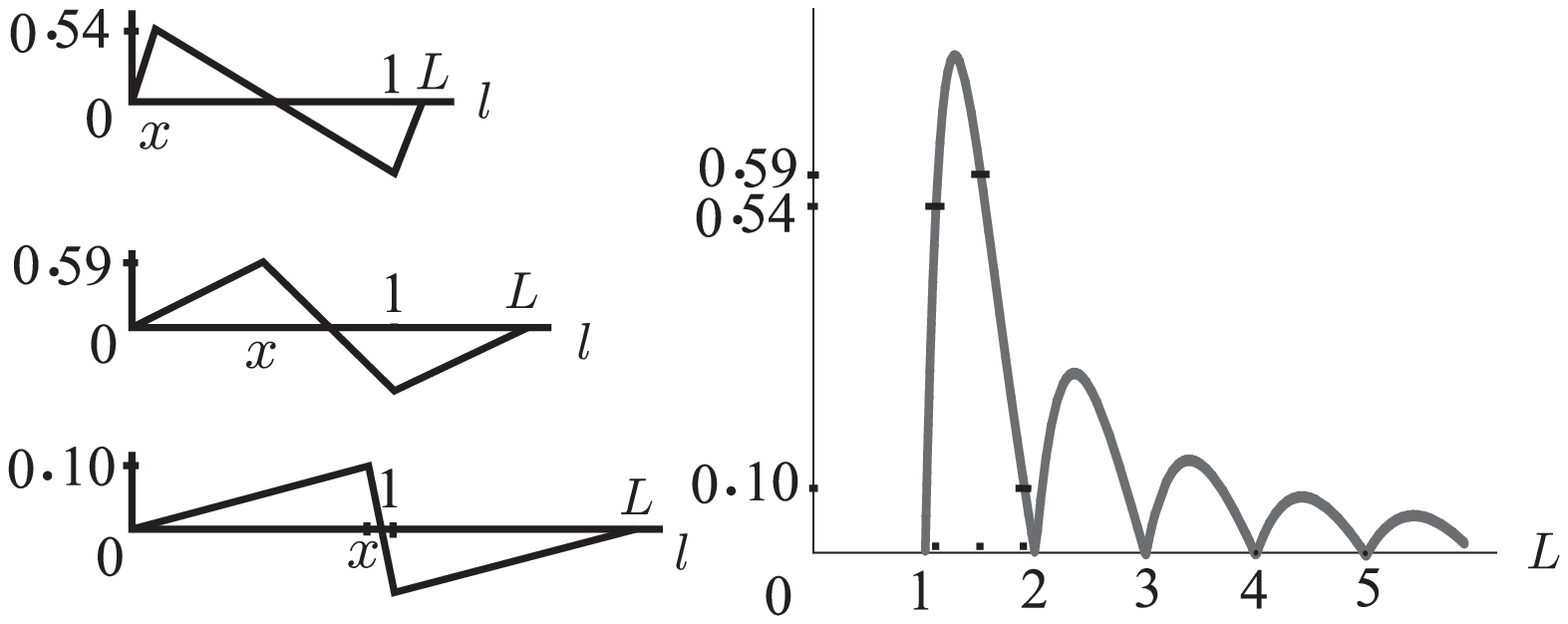}}            
\vskip3mm
      
\noindent {\small {\bf Figure 10} Geometro-topological signature 
$\fuLl$ for $L=$1.1 , 1.5, and 1.9 . } 

\noindent {\small {\bf Figure 11} The function $f(L)$, the absolute 
maximum of $\fuLl$ (which occurs at $l=x$); 
three particular values of $L$ are marked, those used in figure 10.}

\section{When $L>2$}  
                                                   
The generalization of the previous results for arbitrary values of $L$ is 
straightforward but lengthy, so we only state the final results in this 
section. See the appendix 3 for details.  
The graph of the uncorrelated signature (\ref{14}) with 
\bea                                                 \label{17}
L=\lambda+x , \hskip5mm \lambda\in{Z_{+}}, \hskip5mm 0<x<1 
\eea 
has the aspect of a slanted saw; see figure 12, drawn for 
$\lambda=5$ and $x=0.2$. 

\vskip3mm
\hskip1cm\scalebox{0.6}{\ig{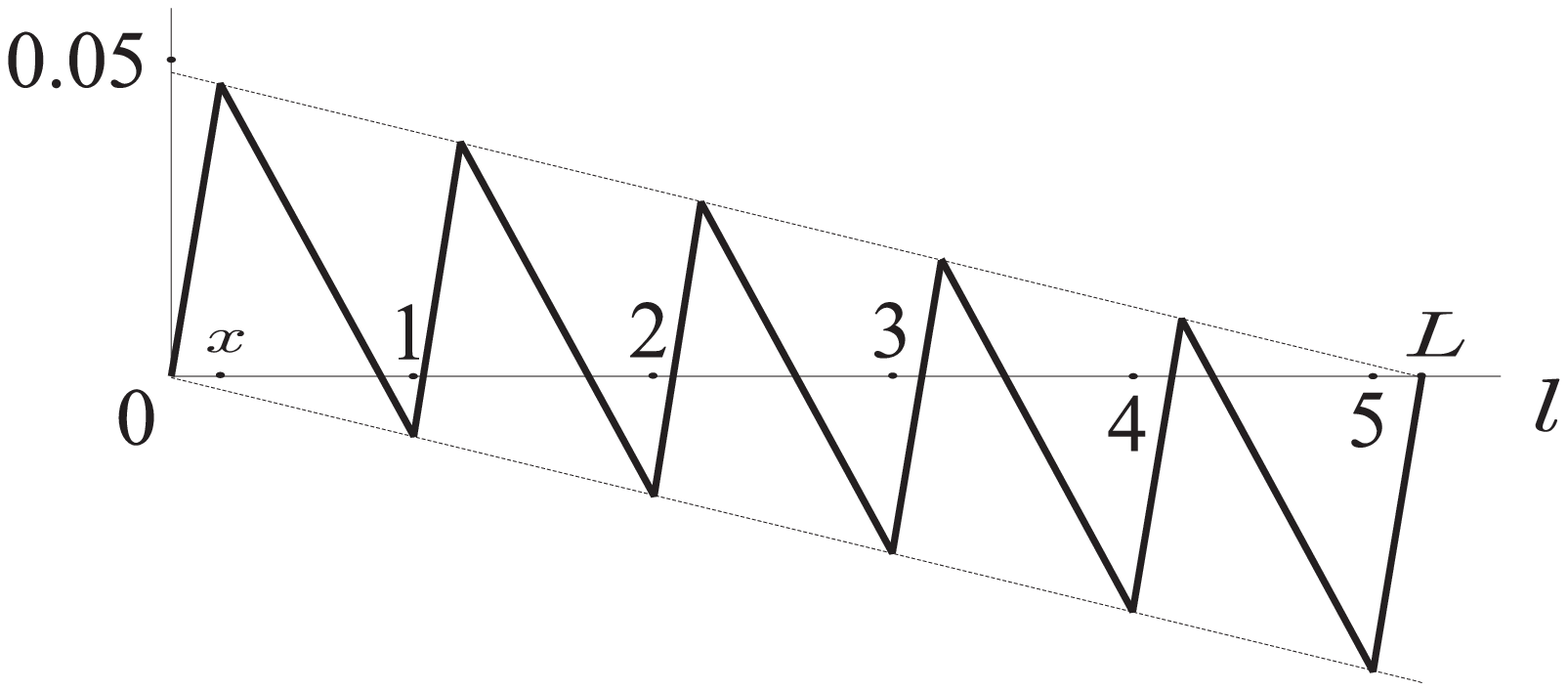}}                     
\vskip3mm 
    
\noindent {\small {\bf Figure 12} The function $\fuLl$ for $L=5.2$ . } 

\vspace{5mm}
\noindent
There are  $\lambda$ maxima, which  occur in the positions 
$l=x, 1+x, ..., (\lambda-1)+x$ , 
and there are $\lambda$ minima, which  lay in the positions 
$l=1, 2,..., \lambda$. 
A straight line connects the maxima, another one connects the minima, 
both have angular coefficient $-8xy/L^3$. 
The $\lambda+1$ segments with positive angular coefficient are parallel, 
as well as the $\lambda$ segments with negative slope. 
As expected, the value of $L$ is the sufficient datum to draw $\fuLl$, 
since 
\bea                                                        \label{18} 
\varphi^u_L(x)=-\varphi^u_L(\lambda)=\frac{8\lambda xy}{L^3}, 
\eea 
as shown in the appendix 4. 
The graph of $f(L)=8\lambda xy/L^3$ is given in Figure 11. 

\section{Conclusion} 

In our first contact with the cosmic crystallography it appeared 
plausible that the normalized expected functions 
$\Fuel$ and $\Fsel$ were the same, since both are concerned with 
separations between objects isometrically unrelated \cite{spikes1}. 
However, in our computer simulations a persistent non-nullity 
of the difference $<\Ful>-\Fsel$ made imperative a more close exam. 
It soon became evident that a difference indeed existed, and that it 
diminished as the number $n$ of objects present in the sample increased. 

Further investigation suggested to define the uncorrelated 
signature \cite{v3} 
\bea                                                         \label{19}
\fuel=(n-1-\sum\nu_g) \Bigl[\Fuel -\Fsel\Bigr] , 
\eea
where $\nu_g=N_g/n$ , with $N_g=$number of $g$-pairs in the observed 
universe; for the cosmic crystallography we usually have 
$n>>1+\sum\nu_g$.

Earlier attempts to find $\Fuel$ for three-dimensional balls failed, 
and also for $2D$ balls; we then focussed our attention on a $1D$ 
ball, this report. When we compare the final theoretical result (\ref{14}) 
with the mean of an increasing number of histograms obtained from
computer simulations, we note a rapid agreement of the two approaches 
in the region  of large separations $l>L/2$, while in the region 
where $l<L/2$ a quite larger number of simulated catalogs is demanded. 
This can be seen in Figure 2, where we observe that the statistical 
fluctuations for $l$ large are sensibly less pronounced than those 
for small $l$. 

When $L<1$, then there is no replication of objects; in this case 
$\FuLl=\FsLl$ and clearly $\fuLl=0$. When $L>1$ is an integer, then 
objects are replicated; nevertheless still $\FuLl=\FsLl$ and $\fuLl=0$. 
This can be seen in the figure 11, where we note that $f(L)$ vanishes 
for $L=$ integer $>0$. 

\vskip5mm {\bf \large Appendix 1}                          
\vskip5mm 
We evaluate the quantity $n-1-\sum\nu_g$ for a universe $S^1$ with 
circumference 1 and observed universe with total amplitude 
$L=\lambda+x$, being $\lambda$ a positive integer and $0<x<1$. 

Assuming $m$ objects along the circle $S^1$ with radius $1/(2\pi)$, 
then the expected number of objects in $L$ is $n=mL$. 
The sum $\sum\nu_g=\nu_{-\lambda}+\nu_{-\lambda+1}+ ... +\nu_{-1}+
\nu_1+ ... + \nu_{\lambda-1}+\nu_\lambda$ indeed simplifies to 
$2(\nu_1+\nu_2+ ... +\nu_\lambda)$, since $\nu_{-i}=\nu_i$. 

Now remember that for $i$ a positive integer $n\nu_i$ is the expected 
number of pairs of objects in the observed universe whose separation 
is $i\leq\lambda$ \cite{spikes1}; its value is 
\bea                                                         \label{20}
n\nu_i=m(L-i) .
\eea
As a consequence $\sum\nu_g=\lambda(L-y)/L$, and finally 
\bea                                                          \label{21}
n-1-\sum\nu_g=(m-1)L-\frac{xy}{L} .
\eea
\vskip5mm {\bf \large Appendix 2}                          
\vskip5mm
We show that $\fuLx=8xy/L^3$ when 
$1<L<2$: from (\ref{11}) or (\ref{12}) we have at $l=1$ 
\bea                                                          \label{22}
\FumpL1=\frac{1}{\Dmp}\frac{p(p-1)}{x} ,
\eea
so we have from (\ref{13})  
\bea                                                         \label{23}
\FumL1=\frac{1}{x}\sum_{p=0}^m \frac{\Pmpx}{\Dmp}p(p-1) , 
\eea
whose value is sought, correct to order $m^{-1}$ when $m>>1$.  
In this limit we have 
\bea                                                          \label{24}
\sum_{p=0}^m \Pmpx (p/m)^k = x^k + 
\frac{k(k-1)}{2m}y\,x^{k-1} +  O(m^{-2}) , 
\eea
and consequently 
\bea                                                         \label{25}
\sum_{p=0}^m \Pmpx F(p/m) = F(x) + 
\frac{xy}{2m}\frac{d^2}{dx^2}F(x) + O(m^{-2}) . 
\eea
For $m>>1$ in eq.(\ref{1}) we find that 
\bea                                                         \label{26}
\frac{p(p-1)}{\Dmp}= \frac{2{\xi}^2}{(\xi+1)^2} + 
\frac{2\xi(2\xi+1)(\xi-1)}{m(\xi+1)^4} + O(m^{-2}) , \hskip5mm 
\xi:=p/m , 
\eea
so from (\ref{25}) we obtain 
\bea                                                         \label{27}
\sum_{p=0}^m \Pmpx\frac{p(p-1)}{\Dmp}&=&\frac{2x^2}{(1+x)^2} + 
\frac{2x(2x+1)(x-1)}{m(1+x)^4} + 
\frac{xy}{2m}\frac{d^2}{dx^2}\Bigl[\frac{2x^2}{(1+x)^2}\Bigr]+O(m^{-2}) 
\nonumber  \\ 
&=& \frac{2x^2}{L^2} - \frac{8x^2 y}{mL^4} + O(m^{-2}) .
\eea
Since $\FsL1=2x/L^2$, we finally have from (\ref{14}), (\ref{23}), 
and (\ref{27}) 
\bea                                                         \label{28}
\fuL1=-\frac{8xy}{L^3} \hskip5mm (1<L<2) . 
\eea
\vskip5mm {\bf \large Appendix 3}                          
\vskip5mm 
We generalize for arbitrary $L>1$ the results obtained for $1<L<2$, 
in particular the equations (\ref{1}) and (\ref{16}). 
We first decompose the total interval $(0, L)$ into $2\lambda+1$ 
subintervals according to figure 13, drawn for $\lambda=5$. 

\vskip3mm
\hskip3cm\scalebox{0.55}{\ig{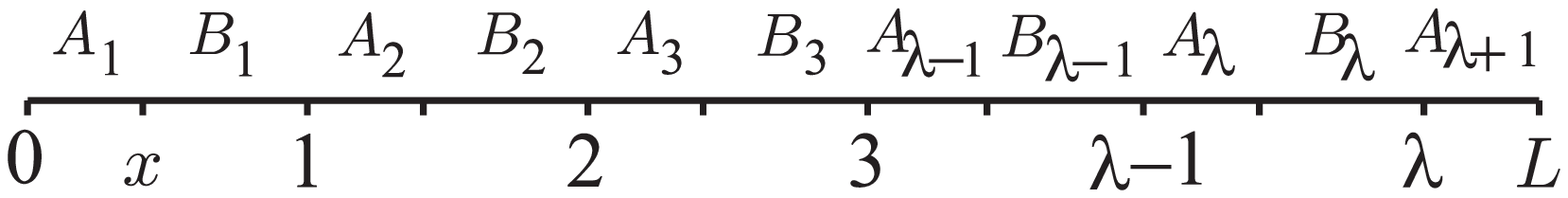}}                       
\vskip3mm     
                                          
\noindent {\small {\bf Figure 13} The one-dimensional observed universe 
with length $L=\lambda+x$, partitioned into $\lambda+1$ intervals $A_i$ 
with length $x$ and $\lambda$ intervals $B_i$ measuring $y=1-x$.} 
\vskip5mm
For $m$ objects randomly distributed in the universe $(0, 1)$ we expect  
$p=mx$ objects in each interval $A_i$ and $m-p=my$ objects in each $B_i$. 
The number of objects in the observed universe $(0, L)$ being 
$m\lambda+p$, the total number of pairs of objects in it is 
$(m\lambda+p)(m\lambda+p-1)/2$; if we deduct the $p\lambda(\lambda+1)/2$ 
correlated pairs with members in the $A$'s, and the 
$(m-p)(\lambda-1)\lambda/2$ correlated pairs with members in the $B$'s, 
then we obtain the expected number of uncorrelated separations 
(cf eq.(\ref{1})): 
\bea                                                         \label{29}
\Dmp=\frac{1}{2}(m\lambda+p)(m\lambda+p-1)-
\frac{1}{2}\lambda(\lambda+1)p - \frac{1}{2}\lambda(\lambda-1)(m-p). 
\eea

We next note in $(0, L)$ the existence of 
\begin{itemize}
\item $w_{A_iA_i}=p(p-1)/2$ pairs with both members in $A_i$;
\item $w_{B_iB_i}=(m-p)(m-p-1)/2$ pairs with both members in $B_i$; 
\item $w_{A_iA_{j>i}}=2w_{A_iA_i}$ uncorrelated pairs, 
      with a member in $A_i$ and the other in $A_{j>i}$; 
\item $w_{B_iB_{j>i}}=2w_{B_iB_i}$ uncorrelated pairs, 
      with a member in $B_i$ and the other in $B_{j>i}$; 
\item $w_{A_iB_{j\geq i}}=p(m-p)$ pairs, with a member in $A_i$ 
       and the other in $B_{j\geq i}$; 
\item $w_{B_iA_{j>i}}=w_{A_iB_{j\geq i}}$ pairs, with a member in $B_i$ 
       and the other in $A_{j>i}$. 
\end{itemize} 
There is a total of $(2\lambda+1)(\lambda+1)$ such numbers $w_{XY}$, and 
their sum  clearly is the $\Dmp$ given in (29). 

With the probability densities $\FXYl$ defined as before, the normalized 
probability density $\FumpLl$ is written similarly to eq. (8), 
\bea                                                           \label{30}
\FumpLl=\frac{1}{\Dmp}\sum_{X,Y}w_{XY}\FXYl . 
\eea
As a matter of fact, there are only three essentially different $w_{XY}$, 
which we dub $w_{AA}, w_{AB}$, and $w_{BB}$, as in sec. 2. 
Also, there are indeed only $3\lambda+1$ different functions 
$\FdXYl$, each appearing with variable multiplicity $m_{XY}$. 
These functions, together with the corresponding $m_{XY}$ and weights 
$w_{XY}$, are displayed in figure 14, drawn for $L=5.2$.

\vskip3mm 
\hskip2cm\scalebox{0.45}{\ig{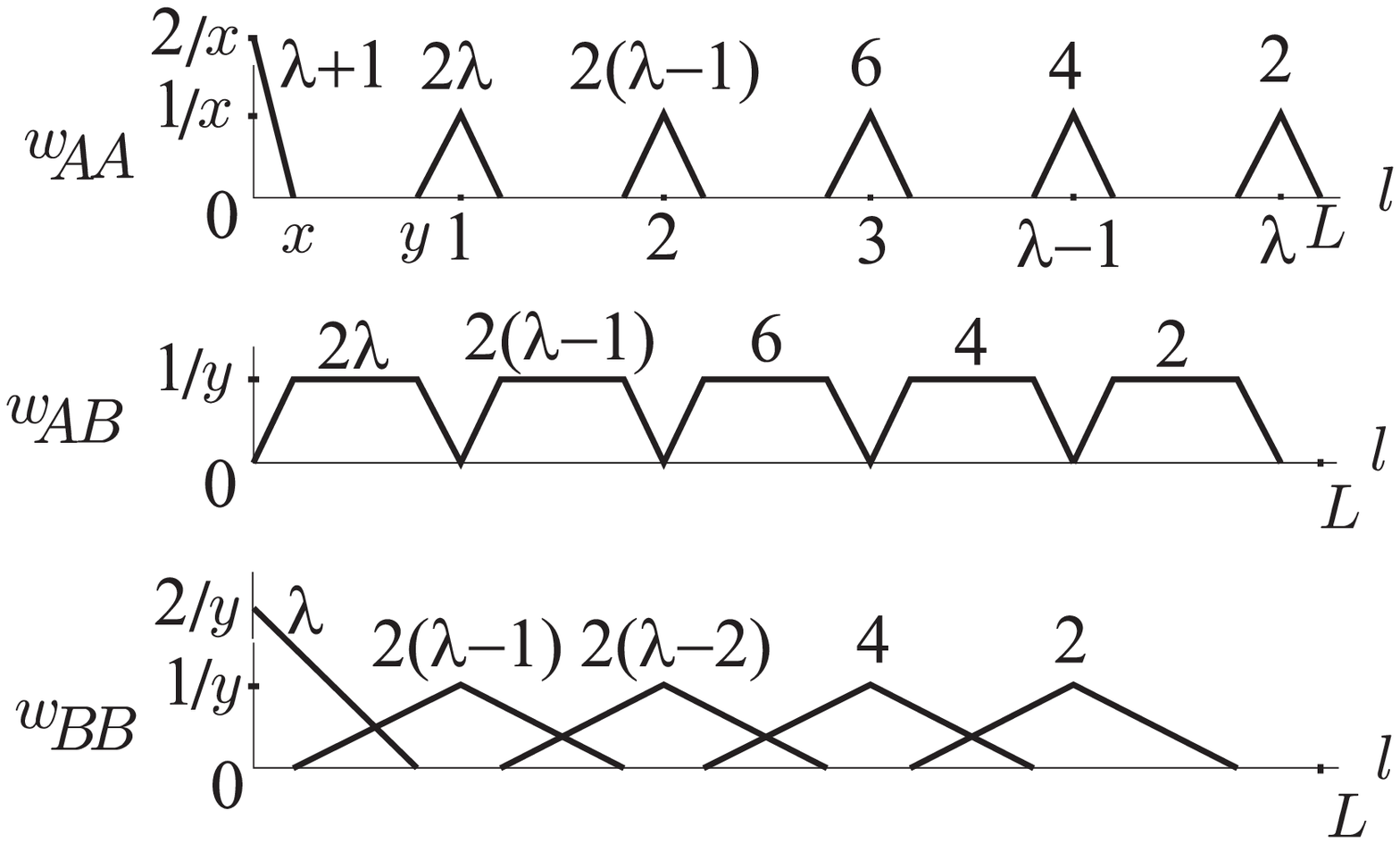}}                         
\vskip3mm     
                                          
\noindent {\small {\bf Figure 14} The $3\lambda+1$ different functions 
$\FdXYl$ when $x\leq 0.5$. On top of each function the corresponding 
multiplicity $m_{XY}$ is written. On the left side the corresponding weight  
$w_{XY}$ is also given. The value $L=5.2$ was taken for definiteness.}
 
\vskip5mm
When $x>0.5$ the set of functions $\FdXYl$ has a different aspect; 
see figure 15, drawn for $L=5.8$. 

\vskip3mm
\hskip2cm\scalebox{0.45}{\ig{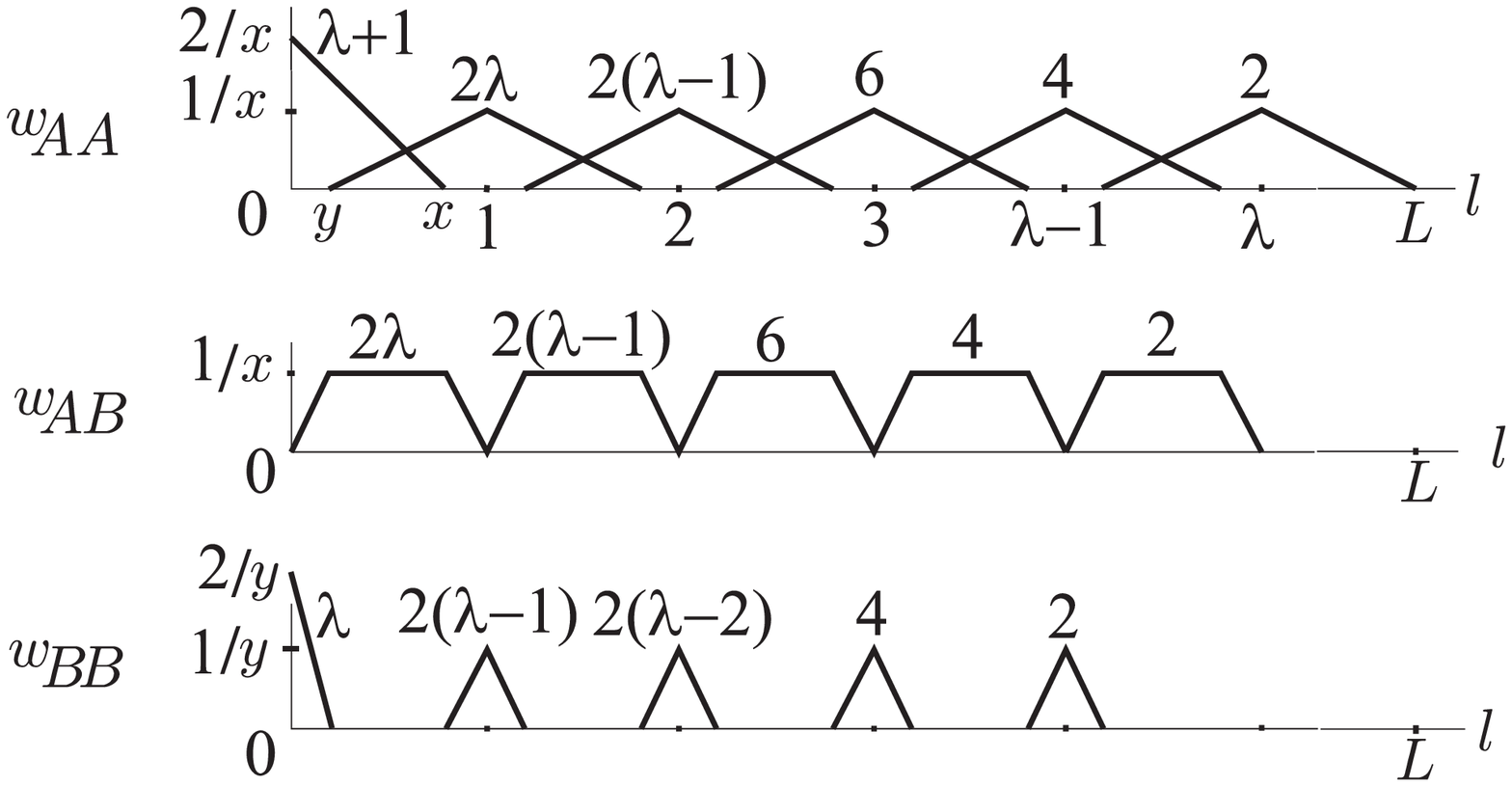}}                         
\vskip3mm     
                                          
\noindent {\small {\bf Figure 15} The $3\lambda+1$ different functions 
$\FdXYl$ when $x\geq 0.5$. The multiplicities $m_{XY}$ and weights   
$w_{XY}$ are indicated as in figure 14. The value $L=5.8$ was taken 
for definiteness.}
 
\vskip5mm
It is now clear that the functions $\FumpLl$ are a sequence of $3\lambda+1$ 
segments, each segment having endpoints either at an integer or separated 
$x$ from an integer; 
as a consequence, also the functions $\FumLl$ (eq.(\ref{13})) have that 
behavior, as well as the functions $\fumLl$ (eq.(\ref{2})). See figure 16. 

\vskip3mm
\hskip3cm\scalebox{0.6}{\includegraphics{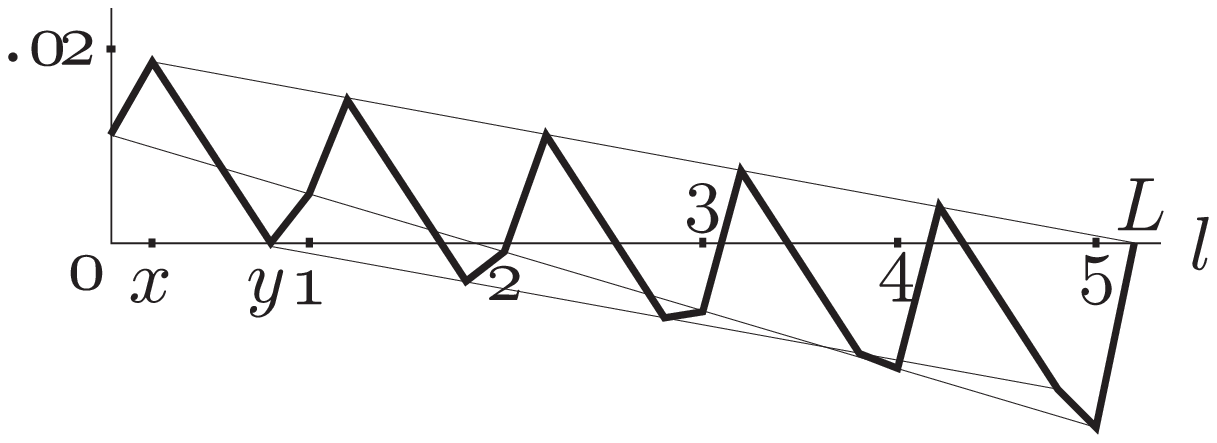}}               
\vskip3mm     
                                          
\noindent {\small {\bf Figure 16} The function $\fumLl$ for $m=2$ and $L=5.2$. 
A straight line connects the points with abscissa $l=i+x \,(i=0,...,\lambda)$; 
another, parallel, connects those with $l=i+y \,(i=0,...,\lambda-1)$; 
also the points with $l=$integer are aligned.

\vskip5mm {\bf \large Appendix 4}           		          
\vskip5mm 
We generalize eq.(\ref{28}) for arbitrary $L>1$. 
For $m>>1$ and $\Dmp$ as in eq.(\ref{29}) we have 
\bea                                                            \label{31}
\frac{p(p-1)}{\Dmp}= \frac{2{\xi}^2}{(\xi+\lambda)^2} + 
\frac{2\lambda\xi(2\xi+\lambda)(\xi-1)}{m(\xi+\lambda)^4} + O(m^{-2}) ,  
\eea
while eq.(\ref{27}) now reads 
\bea                                                             \label{32}
\phi^u_{mL}(\lambda)= \frac{2x^2}{L^2} - \frac{8\lambda x^2 y}{mL^4} 
+ O(m^{-2}) .
\eea
Finally (\ref{28}) becomes 
\bea                                                             \label{33}
\varphi^u_L(\lambda)=-\frac{8\lambda xy}{L^3} , \hskip5mm L>1 . 
\eea
The graph of $f(L)=8\lambda xy/L^3$ is given in figure 11.

\section*{Acknowledgments} 
Conversations with Armando Bernui, Germ\'{a}n I. Gomero, and Marcelo J. 
Rebou\c{c}as are hearty acknowledged.

\end{document}